\documentclass[reprint,aps,prl,superscriptaddress,amsmath,amssymb,floatfix,footinbib,longbibliography]{revtex4-1}
\usepackage{graphicx}
\usepackage{dcolumn}
\usepackage{bm}
\usepackage{amsmath}
\usepackage{times}
\usepackage{color}
\usepackage[breaklinks=true,colorlinks,citecolor=blue,linkcolor=blue,urlcolor=blue]{hyperref}

\makeatletter

\newcommand{\Rmnum}[1]{\expandafter\@slowromancap\romannumeral #1@}
\makeatother

\begin{document}

\title{Common topological origin of longitudinal and transverse magnetoresistance in Fe$_3$GeTe$_2$}
\author{Alapan Bera}
\affiliation{Department of Physics, Indian Institute of Technology Kanpur, Kanpur 208016, India}
\author{Soumik Mukhopadhyay}
\email{soumikm@iitk.ac.in}
\affiliation{Department of Physics, Indian Institute of Technology Kanpur, Kanpur 208016, India}

\begin{abstract} 
In this work, we reveal the coexistence and correlation of a topological cusp anomaly in the planar Hall signal and a spin-flip scattering dominated positive longitudinal magnetoresistance (MR) across the entire temperature range below the Curie point in Fe$_3$GeTe$_2$. This correlation dies out exponentially as the magnetic field is directed away from the $ab$-plane, resulting in unusually sharp polar angle dependence of longitudinal MR and the Hall response.
\end{abstract}

\maketitle
{\it Introduction.---} 
Magnetism in two-dimensional (2D) van der Waals (vdW) systems has become a rapidly growing area of research due to its rich fundamental physics and promising applications in spintronics. The discovery of Fe-based vdW ferromagnets (FMs) has demonstrated robust ferromagnetism down to the monolayer limit~\cite{Roemer2020, Wang2024}, owing to strong perpendicular magnetic anisotropy (PMA). Fe$_\mathrm{n}$GeTe$_2$ (n = 3, 4, 5) is a family of vdW FMs known for their relatively high Curie temperatures (T$_\mathrm{C}$). Among them, Fe$_4$GeTe$_2$ ($\mathrm{T_C}$ $\sim$ 270 K)~\cite{doi:10.1126/sciadv.aay8912, bera2025} and Fe$_5$GeTe$_2$ ($\mathrm{T_C}$ $\sim$ 310 K)~\cite{doi:10.1021/acsnano.8b09660, PhysRevMaterials.3.104401, PhysRevB.110.224401} exhibit (near) room-temperature ferromagnetism. Notably, chemical substitution has enabled the enhancement in Curie point of Fe$_5$GeTe$_2$ well above room temperature~\cite{PhysRevLett.128.217203, doi:10.1126/sciadv.abm7103}. Although Fe$_3$GeTe$_2$ (F3GT) has a comparatively lower $\mathrm{T_C}$ of approx. 230 K~\cite{doi:10.1021/acsnano.2c00512, doi:10.1126/sciadv.aao6791, doi:10.1021/acs.cgd.1c00684}, it exhibits one of the strongest PMA~\cite{Tan2018} among vdW FMs, making it fundamentally useful for spintronic applications. Furthermore, a significantly enhanced Curie temperature of F3GT can be engineered through approaches such as chemical doping~\cite{https://doi.org/10.1002/adfm.202210578}, electrostatic gating~\cite{Deng2018}, and strain engineering~\cite{Hu2020}.

In recent years, significant efforts have been made to understand the magneto-transport properties of F3GT, both in bulk and few-layer devices. In addition to the intrinsic momentum-space Berry phase-driven anomalous Hall effect (AHE)~\cite{Kim2018, PhysRevB.96.134428, PhysRevB.107.035115} for the conventional orthogonal configuration, F3GT also exhibits exotic features when the magnetic field and electric field (current) are oriented along the basal plane, irrespective of their internal azimuthal angle. There have been reports of a topological Hall effect (THE)~\cite{PhysRevB.96.134428, PhysRevB.100.134441} in bulk F3GT, identified by a cusp anomaly in the Hall resistivity, which is associated with a non-zero scalar spin chirality arising from the real-space Berry curvature induced by a non-coplanar spin texture. Few recent studies of magnetoresistance (MR) in a planar field-current configuration in bulk~\cite{PhysRevResearch.6.L032008} and nano-flake~\cite{Zeng_2022} F3GT revealed a spin-flip scattering dominated non-trivial positive MR behavior in the unsaturated magnetic phase, which scales quadratically to the magnetic field and eventually transits to a trivial negative MR in the saturated regime. While the THE and spin-flip scattering dominated positive MR in F3GT have been demonstrated separately in various studies, a coexistence and correlation between these phenomena remains lacking. In those studies where the THE is reported~\cite{PhysRevB.96.134428, PhysRevB.100.134441}, the absence of positive MR is most likely attributed to a sample imperfection due to vacant Fe sites. This also explains the lower Curie temperature ($\mathrm{T_C}$) than the ideal $\mathrm{T_C}$ = 230 K, along with comparatively lower PMA strength, which is a key to the positive planar MR. Whereas in the other works~\cite{Zeng_2022, PhysRevResearch.6.L032008} where the positive MR is recorded, either it has been studied in nano-flake F3GT, where THE is not observed, or they did not explore Hall resistivity in the planar configuration.

In this report, with a detailed study of magneto-transport and magnetization in bulk F3GT at various temperatures and magnetic field orientations, we demonstrate the coexistence and correlation of spin-flip scattering dominated positive MR with the topological Hall signature. This correlation persists across the entire temperature range below $\mathrm{T_C}$, as well as within the field range below the saturation point of $ab$-plane magnetization. Additionally, we present the evolution of these features with the out-of-plane field-angle rotation to point out a sharp anisotropic MR behaviour attributed to the dominant spin-flip scattering confined to a narrow alignment window.

{\it Experimental methods.---} Single crystals of F3GT are grown using the chemical vapor transport (CVT) technique~\cite{doi:10.7566/JPSJ.82.124711} with iodine (I$_2$) as the transport agent. The average crystal dimension is approximately 1 mm. Single-crystal X-ray diffraction (XRD) is performed using a Panalytical XRD setup to confirm the high crystallinity of the crystals, whereas energy-dispersive X-ray spectroscopy (EDS) measurement (using JEOL W-SEM) is carried out on the cleaved surfaces to verify the elemental composition. Temperature and magnetic field dependence of DC magnetization are measured using a Physical Property Measurement System (PPMS) manufactured by Oxford Instruments. Temperature and field dependence of electrical resistivity are studied using a standard lock-in technique with a 12 T Variable Temperature Insert (VTI) provided by Cryogenics Ltd. Magnetoresistivity and Hall resistivity are measured simultaneously using a six-probe contact configuration and are symmetrized and antisymmetrized, respectively, to eliminate the effects of probe misalignment.

 \begin{figure}[t]
\includegraphics[width=\linewidth]{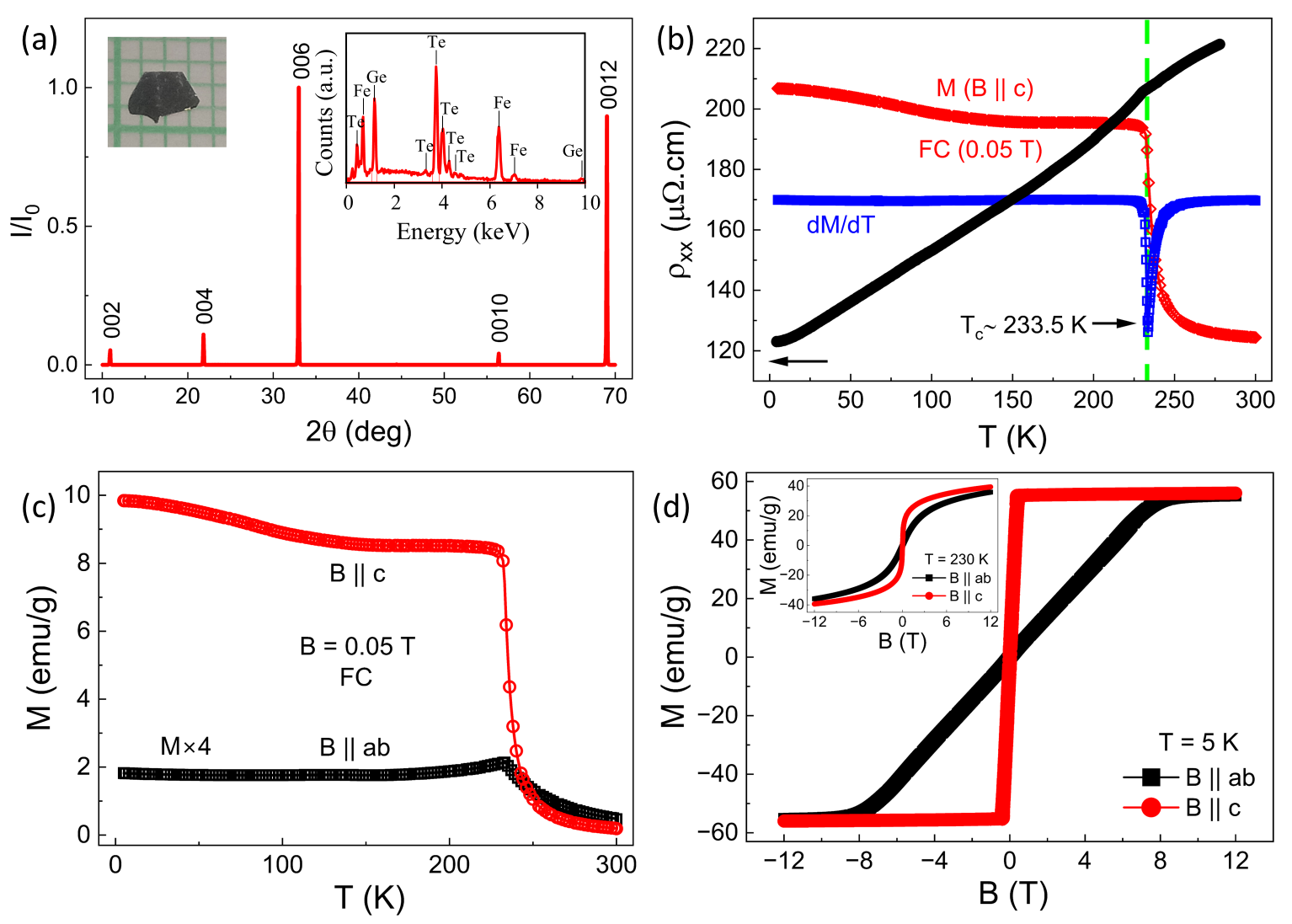}
\caption{(a) Normalized (00l) XRD peaks confirm the single-crystalline nature of bulk F3GT. In the inset (right), the EDS spectra show the prominent Fe, Ge, and Te peaks. An optical image of a bulk crystal is presented in the left inset. (b) Temperature dependence of resistivity and DC magnetization (for B$\parallel$c) of bulk F3GT. The FM-PM transition can be seen near 233.5 K from the derivative of the $M-T$ curve. (c) Temperature dependence of the DC magnetization shows a large disparity of the magnetization for B$\parallel$ab and B$\parallel$c, showing the presence of a large PMA. (d) Field dependence of DC magnetization at 5 K (at 230 K in the inset) is presented for the field applied parallel and perpendicular to the $c$-axis. The large saturation field along the $ab$-plane is the manifestation of the strong PMA.}
\label{fig1}
\end{figure}

\begin{figure}[t]
\includegraphics[width=\linewidth]{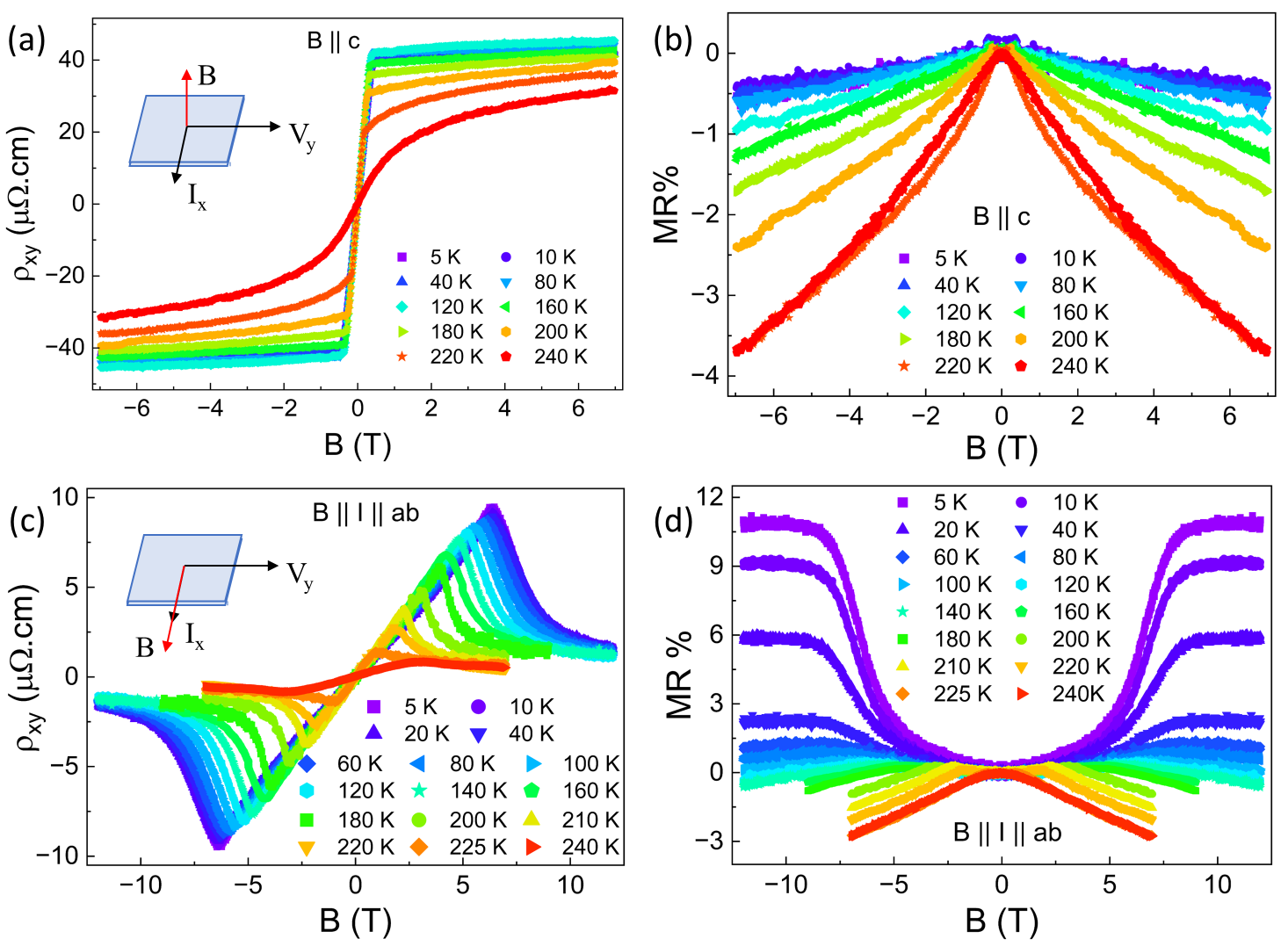}
\caption{(a) Field dependence of Hall resistivity for B$\parallel$c at different temperatures shows the presence of a large AHE below the ordering temperature. (b) MR vs. B curves for B$\parallel$c show a typical negative MR originating from field-induced suppression of electron-magnon scattering. (c) Hump-like signature in the field dependence of Hall-resistivity for B$\parallel$I$\parallel$ab indicates the presence of THE in the entire temperature range below $\mathrm{T_C}$. (d) Field-dependence of longitudinal MR for B$\parallel$ab at different temperatures showing an inflection point at a critical field due to a transition from spin-flip scattering dominated positive MR to a negative MR due to field suppression of electron-magnon scattering. The measurement configurations are presented in the inset of (a) and (c).}
\label{fig2}
\end{figure}

{\it Results and discussion.---} F3GT crystallizes in a hexagonal lattice structure with space group P6$_3$/mmc, where Fe$_3$Ge slabs are sandwiched between Te layers. Fig.~\ref{fig1}(a) presents the single-crystal XRD data taken from a freshly cleaved surface of a bulk F3GT crystal. The prominent (00l) peaks indicate high crystallinity, with the $c$-axis oriented perpendicular to the sample's $ab$-plane. EDS spectra in the inset (right) of Fig.~\ref{fig1}(a) show the pronounced Fe, Ge, and Te peaks with a stoichiometric ratio of $\sim$ 2.87 : 1.05 : 2.10, close to the nominal composition of 3 : 1 : 2. An optical image of a typical F3GT single crystal is presented in the left inset. 

The temperature dependence of the electrical resistivity ($R-T$), shown in Fig.~\ref{fig1}(b), exhibits metallic behavior, with a slope change near 233 K indicating a ferromagnet-paramagnet (FM-PM) phase transition. This transition is also observed in the temperature dependence of the field-cooled magnetization ($M-T$) curve (more prominent in its temperature derivative) measured at magnetic field $\mathrm{B} = 0.05$ T applied along the $c$-axis. As presented in Fig.~\ref{fig1}(c), the magnetization with the field applied along the $c$-axis (red symbols) is significantly larger than that for the $ab$-plane (black symbols), indicating the presence of strong PMA in F3GT. The field dependence of the DC magnetization ($M–H$) curves for $ab$-plane and $c$-axis orientations, shown in Fig.~\ref{fig1}(d), reveals a large difference in saturation fields at 5 K, due to the strong PMA. This anisotropy remains significant even near $\mathrm{T_C}$, as evident from the $M–H$ curve at 230 K (inset), establishing the $c$-axis as the easy axis throughout the ordered phase.

Next, we investigate the magneto-transport properties of F3GT by examining the field dependence of the magnetoresistivity ($\rho_{\mathrm{xx}}$) and Hall resistivity ($\rho_{\mathrm{xy}}$) in the temperature range of 5--240 K. Fig.~\ref{fig2}(a) presents the Hall resistivity curves for B$\parallel$c at various temperatures. At all temperatures below $\mathrm{T_C}$, a sharp increase in $\rho_{\mathrm{xy}}$ is observed in the low-field regime, attributed to the anomalous Hall effect (AHE) arising due to the ferromagnetic nature of F3GT. In contrast, at 240 K, this sharp AHE signature is absent, consistent with the transition to a paramagnetic phase. Fig.~\ref{fig2}(b) shows the MR curves for the field applied along the easy $c$-axis. A typical negative MR, resulting from the field-induced suppression of electron-magnon scattering, is observed across the entire temperature range as seen in conventional ferromagnets. As the temperature increases, the enhanced magnon excitation leads to a greater percentage change in the negative MR.

For the field applied along the crystal $ab$-plane, F3GT exhibits intriguing magneto-transport phenomena. Fig.~\ref{fig2}(c) shows the Hall resistivity curves over a wide temperature range for B$\parallel$I$\parallel$ab configuration. A distinct cusp-like feature is observed at all temperatures below $\mathrm{T_C}$, indicating a pronounced THE. While the most common origin of THE is the presence of topological spin textures such as skyrmions or merons, non-collinear and non-coplanar spin arrangements can also give rise to a non-zero scalar spin chirality. Conduction electrons traversing these spin textures acquire a finite Berry phase in real space, which leads to a fictitious magnetic field, causing an additional contribution to the Hall resistivity. The spins are forced into a collinear configuration at sufficiently high magnetic fields, causing THE contribution to die out, resulting in the observed cusp-like feature. The THE persists throughout the entire temperature range below $\mathrm{T_C}$, with the magnitude of the cusp decreasing monotonically as the temperature rises.

\begin{figure}[t]
\includegraphics[width=0.7\linewidth]{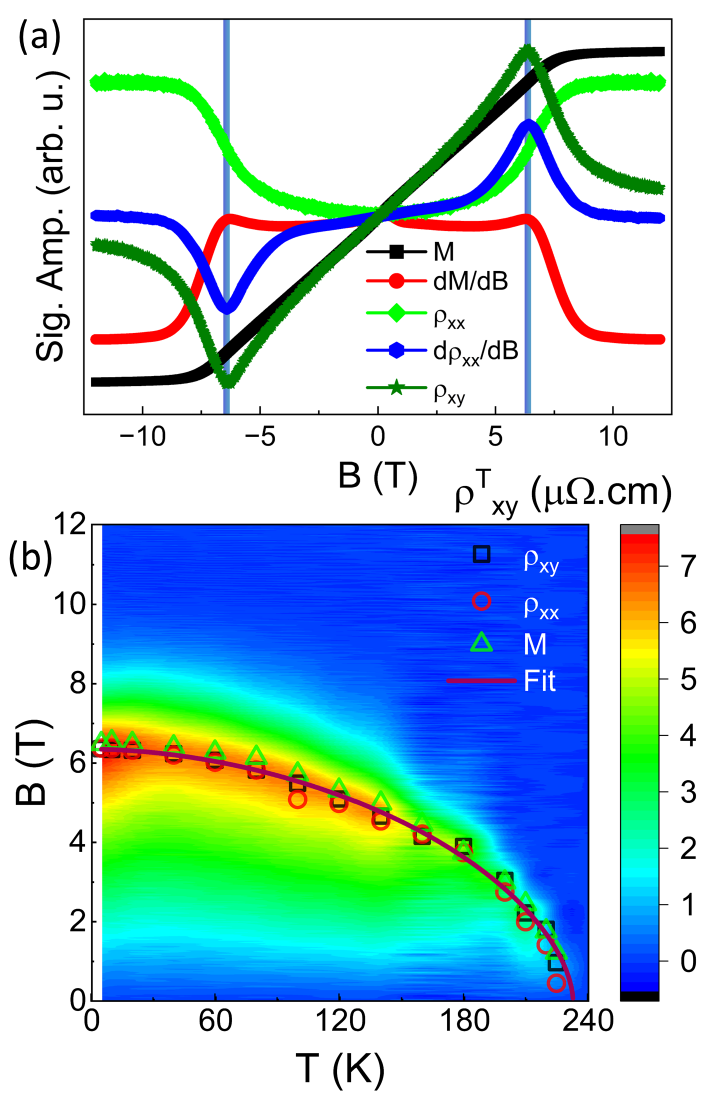}
\caption{(a) Correlation between the critical field points is depicted by the violet line that connects the change of curvature in the MR, Hall, and $M-H$ curves. Peak in the field derivative of the MR curve (blue symbols) points to the onset of change in the MR mechanism. The y-axis is not to scale. (b) Contour plot of THE contribution to the $ab$-plane Hall curves shows a monotonous increase of the THE amplitude as the temperature decreases. The temperature dependence of the critical fields calculated from the field derivative of the MR curves (symbols in red) is correlated to the onset of saturation of the $ab$-plane magnetization and the field corresponding to the maximum THE amplitude. The red line represents a fit of the temperature evolution of B$^*$ to a function analogous to the temperature dependence of magnetization, $\mathrm{B^*(T)}=\mathrm{B^*(0)\left(1-\left(T / T_{c}\right)^\alpha\right)^\beta}$, with $\alpha= 1.83$ and $\beta= 0.57$.}
\label{fig3}
\end{figure}

The intrinsic Berry phase-driven Karplus-Luttinger (K-L) mechanism is the primary contributor to the AHE signal in F3GT~\cite{Kim2018, PhysRevB.96.134428, PhysRevB.107.035115}. Thus, the total Hall resistivity, $\rho_{\mathrm{xy}}$, can be expressed as
\begin{equation}
\rho_{\mathrm{xy}}=\rho_{\mathrm{xy}}^\mathrm{O}+\rho_{\mathrm{xy}}^\mathrm{A}+\rho_{\mathrm{xy}}^\mathrm{T}=\mathrm{R}_0 \mathrm{B}+\mathrm{S_H \rho^2_{xx}M}+\rho_{\mathrm{xy}}^\mathrm{T}~\label{Eq1}
\end{equation}
where $\rho_{\mathrm{xy}}^\mathrm{O}$, $\rho_{\mathrm{xy}}^\mathrm{A}$, and $\rho_{\mathrm{xy}}^\mathrm{T}$ represent the ordinary, anomalous, and topological Hall contributions, respectively. Here, we have used the quadratic scaling relation between the anomalous Hall coefficient $\mathrm{R_S}$ and the electric resistivity $\rho_{\mathrm{xx}}$, in accordance with the K-L theory: $\mathrm{R_S}= \mathrm{S_H \rho^2_{xx}}$, where $\mathrm{S_H}$ is the anomalous Hall factor. $\mathrm{R_0}$ and $\mathrm{M}$ denote the ordinary Hall coefficient and saturation magnetization, respectively. The ordinary and anomalous Hall contributions can be separated through a scaling analysis, using the in-plane field-dependent magnetization and MR curves. As $\rho_{\mathrm{xy}}^\mathrm{T}$ becomes negligible in the high-field saturation regime due to the coplanar and collinear spin arrangement, $\mathrm{R_0}$ and $\mathrm{S_H}$ in Eq.~\ref{Eq1} can be determined with the linear fitting of $\mathrm{\rho^2_{xx}M}/H$ vs. $\rho_{\mathrm{xy}}/\mathrm{H}$. The obtained $\mathrm{R_0}$ and $\mathrm{S_H}$ values are then substituted back into the expressions of  $\rho_{\mathrm{xy}}^\mathrm{O}$ and  $\rho_{\mathrm{xy}}^\mathrm{A}$ to calculate the ordinary and anomalous Hall contributions across the entire field range. The THE amplitude is extracted by subtracting these contributions from the measured $\rho_{\mathrm{xy}}$ curve. The contour plot showing the field and temperature dependence of the THE contribution is presented in Fig.~\ref{fig3}(b).

Fig.~\ref{fig2}(d) presents the field dependence of the longitudinal (B$\parallel$I$\parallel$ab) MR in F3GT. Interestingly, in contrast to the typical negative MR for B$\parallel$c, we observe a non-trivial positive MR in the planar configuration, which increases monotonically with the field strength B up to a critical field B$^*$, beyond which a negative MR starts to emerge. This MR behavior arises from the interplay between two scattering mechanisms that influence the resistivity under an applied magnetic field. Below B$^*$, as the field strength increases, more spins begin flipping from the easy $c$-axis toward the $ab$-plane to minimize the Zeeman energy. As a result, spin-flip scattering increases with the field strength, leading to a positive MR. Thus, as expected, B$^*$ corresponds to the saturation point of the magnetization for B$\parallel$ab. Although field-induced suppression of electron-magnon scattering, which typically leads to negative MR in FMs, is also present below B$^*$, spin-flip scattering dominates in this field regime. Above B$^*$, with all spins aligned along the $ab$-plane, spin-flip events diminish, thus paving the way for the usual negative MR to show up.

\begin{figure}[t]
\includegraphics[width=\linewidth]{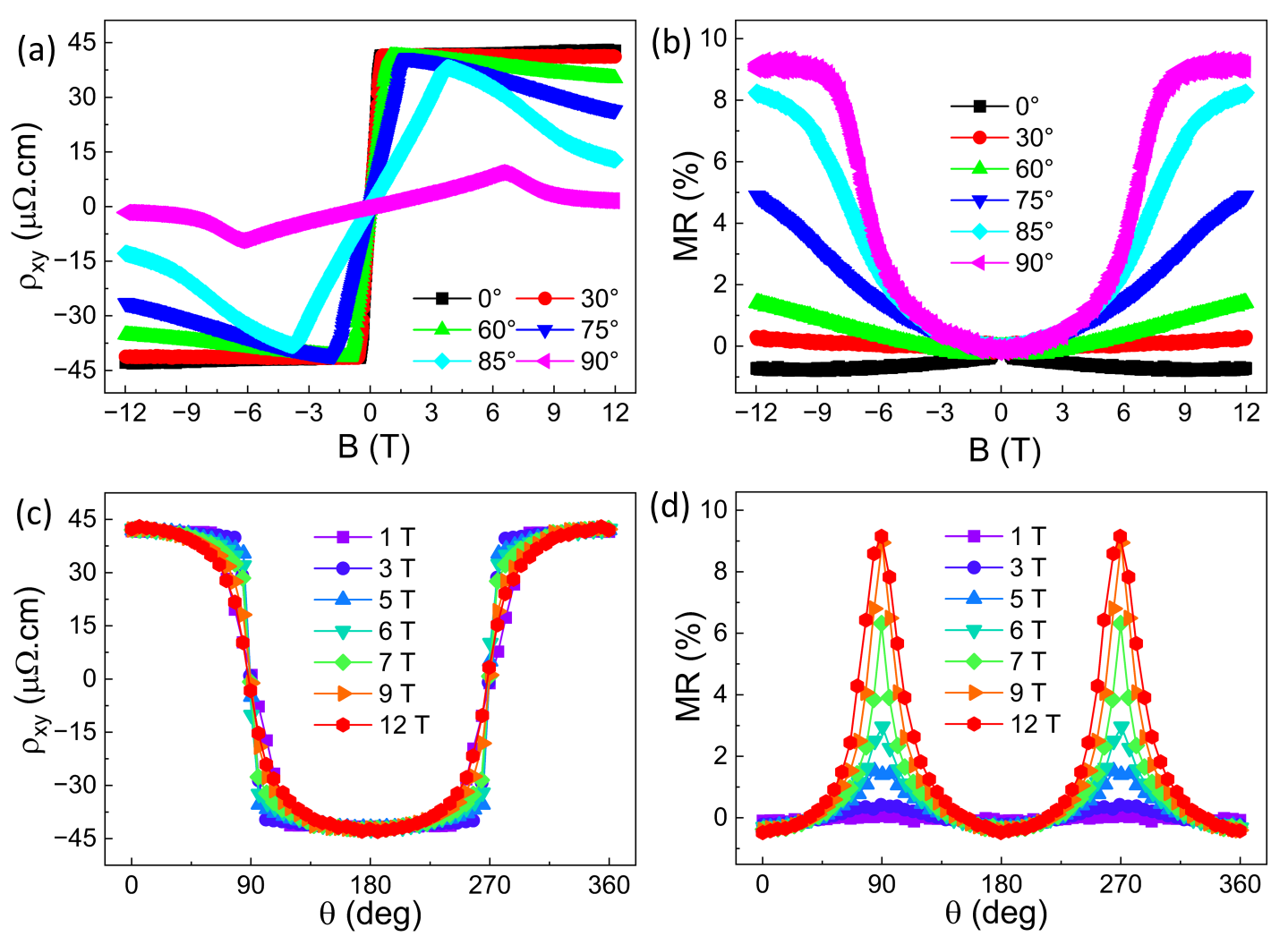}
\caption{Evolution of the (a) Hall and (b) MR curves for field orientation along different polar angles $\theta$ with respect to the sample $ab$-plane at 10 K. The cusp feature in the Hall curves and the slope-change in the MR curve at B$^*$ are visible only in a narrow $\theta$ window around 90$^\circ$. (c) Angle-dependence of Hall resistivity at various constant magnetic fields, showing larger deviation from the typical $\cos\theta$ behavior at lower field strengths. (d) A sharp rise in the MR is observed at around $\theta=90^\circ$ due to the strong spin-flip scattering at the non-coplanar spin-arrangement.}
\label{fig4}
\end{figure}

Although observations of the THE were made earlier in bulk F3GT~\cite{PhysRevB.96.134428, PhysRevB.100.134441} in the $ab$-plane configuration of magnetic field and current, no such MR behavior has been reported in those studies. While Jiezun Ke et al.~\cite{Ke_2020} reported a study of MR in bulk F3GT in the planar configuration of field and current, no signature of such non-trivial positive MR was observed.  This is because spin-flip scattering becomes more pronounced when the PMA is significantly stronger than the shape anisotropy associated with the sample geometry. In the case of Ke et al., the relatively small difference between the magnetization for B$\parallel$c and B$\parallel$ab suggests a diminished PMA strength. Fe-vacancy induced lattice imperfection may lead to this effect, which also explains the observation of a significantly lower $\mathrm{T_C}$ of 138 K. In contrast, the PMA strength and Curie temperature of 233 K presented in this study are higher than those observed in most bulk F3GT systems reported to date~\cite{PhysRevB.100.134441, PhysRevB.96.134428, PhysRevB.107.035115, https://doi.org/10.1002/adma.202311831, PhysRevResearch.6.L032008, Liu2017, Chowdhury2021, https://doi.org/10.1002/ejic.200501020}.

Fig.~\ref{fig3}(a) presents the $M–H$, Hall, and MR curves, along with the field derivative of the MR and $M–H$ curves, measured at 5 K for the B$\parallel$I$\parallel$ab configuration. The violet line highlights the critical field B$^*$, which corresponds to a change in slope in the MR curve (also marked by a peak in its field derivative), the field at which the cusp amplitude in the Hall curve is maximized, and the onset of saturation in the magnetization for B$\parallel$ab (more evident from its field derivative). This indicates a direct correlation between spin-flip scattering events and the emergence of the THE. As the temperature increases, the reduction in spin stiffness leads to a lower saturation field for magnetization along the $ab$-plane, consequently lowering the critical fields with rising temperature. Fig.~\ref{fig3}(b) shows the temperature dependence of these correlated critical fields, along with the THE contour plot. This correlation persists throughout the entire temperature range corresponding to the ordered state of F3GT. The temperature evolution of the critical fields is fitted with a scaling function analogous to the temperature dependence of magnetization~\cite{Zhang_2025}:
\begin{equation}
\mathrm{B^*(T)} = \mathrm{B^*(0)\left[1 - \left(T / T_{c}\right)^\alpha\right]^\beta},
\end{equation}
where $\alpha = 1.83$ and $\beta = 0.57$. This confirms that the non-coplanar and unsaturated magnetic phase plays a crucial role in the observed phenomena.

Fig.~\ref{fig4}(a) and ~\ref{fig4}(b) show the Hall and MR curves for the magnetic field oriented along various polar angles $\theta$ within the plane defined by the applied current I and the $z$-axis. An evolution from the AHE to a THE signature is evident in the Hall curves as the field orientation shifts from the conventional orthogonal configuration ($\theta = 0^\circ$) to the planar configuration ($\theta = 90^\circ$). A corresponding transition from negative to positive MR is also observed. Notably, while positive MR begins to appear at $\theta = 30^\circ$, the field-driven transition at B$^*$ from a positive to negative MR becomes pronounced only from $\theta = 85^\circ$ onward, indicating that the effect is confined to a narrow $\theta$ range around $90^\circ$. Fig.~\ref{fig4}(c) presents the $\theta$-dependence of the Hall resistivity at various fixed magnetic field strengths. At low fields, the $\rho_{\mathrm{xy}}$ vs. $\theta$ curves deviate from the typical $\cos \theta$ behavior due to contributions from the THE. However, at higher fields (near 12 T), the curves approach the $\cos \theta$ dependence, as the THE contribution diminishes with the coplanar spin arrangement in the saturated magnetic phase. Fig.~\ref{fig4}(d) presents the anisotropic MR (AMR) curves with the expected two-fold anisotropy. However, instead of the typical $\cos 2\theta$ behavior, a sharp peak in AMR is observed near $\theta$ = $90^\circ$ and $270^\circ$, which is attributed to spin-flip scattering being dominant only within a narrow angular window around the planar configuration.

In summary, using a temperature, magnetic field, and angle-dependent study of magneto-transport, we uncover a correlation between the emergence of non-coplanar spin arrangement-induced THE and spin-flip scattering-driven positive MR in the unsaturated magnetization state of bulk F3GT, in the planar alignment of the magnetic field and current. This correlation persists throughout the entire ordered state of the system and becomes weaker with increasing temperature. Out-of-plane angular measurements reveal that the observed phenomena are narrowly confined in the polar angle space, becoming prominent only around $\theta = 90^\circ$. Our study highlights a relatively unexplored aspect of F3GT in the planar configuration and offers insights into the transport properties in the frustrated, unsaturated phase of a vdW FM with strong PMA. Furthermore, the sharp angular peak features observed in the AMR may have potential applications in magnetic sensing with precise angular sensitivity.

{\it Acknowledgements.---} The authors acknowledge IIT Kanpur and the Department of Science and Technology, India, [Order No. DST/NM/TUE/QM-06/2019 (G)] for financial support. AB thanks PMRF for financial support.

\end{document}